# Current-perpendicular-to-plane giant magnetoresistance of a spin valve using $Co_2MnSi$ Heusler alloy electrodes

K. Kodama[1], T. Furubayashi[2 *], H. Sukegawa[2], T. M. Nakatani[1], K. Inomata[2] and K. Hono[2,1]

[1] *Graduate School of Pure and Applied Sciences, University of Tsukuba, Tsukuba 305-0047, Japan*

[2] *National Institute for Materials Science, Tsukuba 305-0047, Japan*

[*] Electronic mail: furubayashi.takao@nims.go.jp

We report the current-perpendicular-to-plane giant magnetoresistance of a spin valve with $Co_2MnSi$ (CMS) Heusler alloy ferromagnetic electrodes. A multilayer stack of Cr/Ag/Cr/CMS/Cu/CMS/$Fe_{25}Co_{75}$/$Ir_{28}Mn_{72}$/Ru was deposited on a MgO (001) single crystal substrate. The bottom CMS layer was epitaxially grown on the Cr/Ag/Cr buffer layers and was ordered to the $L2_1$ structure after annealing at 673 K. The upper CMS layer was found to grow epitaxially on the Cu spacer layer despite the large lattice mismatch between Cu and CMS. The highest MR ratios of 8.6% and 30.7% for CPP-GMR were recorded at room temperature and 6 K, respectively. The high spin polarization of the epitaxial CMS layers is the most likely origin of the high MR ratio.

Recently, tunneling magnetoresistance (TMR) devices were successfully applied to read heads for high-density hard disc drives (HDDs) over 200 Gbit/in$^2$ areal density because FeCoB/MgO/FeCoB based magnetic tunneling junctions (MTJs) exhibit satisfactory performance such as 50% room temperature TMR ratio with the resistance-area product of 0.4 $\Omega\,(\mu m)^2$ [1]. However, for still higher areal density exceeding 500 Gbit/in$^2$, much lower resistance-area product that is difficult to be achieved in MTJs is required. The intrinsically low resistance of metal-made CPP-GMR devices is considered to have an advantage over TMR devices for achieving high-speed read out. However, the typical MR ratio of CPP-GMR spin valves is only around 2% at room temperature (RT), which is too low for read head applications. Thus, it is crucial to increase the MR ratio of the current CPP-GMR devices substantially for potential applications.

Since the highly spin polarized current is expected to enhance the CPP-GMR ratios,[2] many attempts have been made to achieve high MR ratios utilizing highly spin-polarized Heusler alloys as ferromagnetic electrodes.[3-5] Some of the Heusler alloys with the $L2_1$ ordered structure are predicted to be half-metal based on *ab initio* calculations.[6-8] Recent theoretical work by Tang *et al.* [9] suggested that the MR ratio for the CPP-GMR using the half-metal electrodes could be as high as 60%. However, the experimental work on CPP-GMR using $Co_2MnSi$ (CMS) Heusler alloy by Yakushiji *et al.*[3] reported a MR ratio of only 2.4 % at RT in an epitaxial CMS/Cr/CMS tri-layer pseudo spin valve, although CMS was demonstrated to be a highly spin-polarized half-metal at low temperatures from the TMR measurements by Sakuraba *et al.*[10] More recently, Nikolaev *et al.*[5] reported a value of 9 % at RT; however, the type of the ferromagnetic electrode was not revealed. Thus, the report of high MR values in CPP-GMR using the Heusler alloy electrodes is still very limited. We have recently demonstrated that a relatively large MR ratio of 6.9% at room temperature can be obtained from a spin valve using $Co_2FeAl_{0.5}Si_{0.5}$ Heusler alloy electrodes with B2 order.[11] Thus, we expected that higher MR ratio may be achieved if an electrode material with a higher tendency of $L2_1$ order is selected.

In this paper, we report a large MR ratio from a CPP-GMR spin valve using a $Co_2MnSi$ (CMS) Heusler alloy. We prepared films with a layer structure of Cr/Ag/Cr/CMS/Cu/CMS/$Co_{75}Fe_{25}$/$Ir_{22}Mn_{78}$/Ru from the bottom on MgO(001) single crystal substrates. Note that the Heusler alloy layers were epitaxially grown on MgO (001) substrates with appropriate choices of buffer layers in the successful reports of high TMR values.[10, 12, 13] This is possibly because the ordered structure of $L2_1$ or $B2$, which is necessary for obtaining high spin polarization, can be realized more easily in the epitaxially grown films than in the polycrystalline films. Therefore, we also used epitaxial CMS layers for the fabrication of CPP-GMR spin valves. A Cr buffer layer has been used in TMR junctions using CMS [10, 12] as well as in CPP-GMR films of CMS/Cr/CMS by Yakushiji. *et al.*[3] However, we chose a Cr/Ag/Cr trilayer for the buffer layer to reduce the electrical resistance by the insertion of the Ag layer. The buffer layers serve as the bottom electrode for measuring the CPP-GMR. Thus, this leads to the reduction of the total resistance of the stack. An epitaxial growth of CMS on the buffer layer was still achieved because of the epitaxial relationship of $(001)_{Cr}//(001)_{Ag}$ and $[100]_{Cr}//[110]_{Ag}$ from the lattice constants of $a$ = 0.2884 nm for Cr and $a/\sqrt{2}$ = 0.2888 nm for Ag. The choice of the metallic spacer layer is also important. The $\Delta_1$ band of Cr has an energy gap at the Fermi surface and does not contribute to the conduction.[14] Thus, the Cr spacer layer with the (001) epitaxial structure may have a disadvantage of blocking the spin polarized current from the CMS $\Delta_1$ band. In this work, we adopted Cu for the spacer layer, which shows no such energy gap at the Fermi surface. In addition, Cu is most commonly used as the spacer layer of CPP-GMR devices because of the low resistivity and the long spin diffusion length. Note that the upper CMS layer on the Cu spacer layer also grew epitaxially on the (001) layer despite the large lattice mismatch between Cu ($\sqrt{2}\,a$ = 0.5112 nm) and CMS ($a$ = 0.566 nm). Thus, Cu was found to be a suitable material as the spacer layer.

The multilayers were grown by dc magnetron sputtering in an ultrahigh vacuum system with a base pressure below $5.0\times10^{-7}$ Pa. The target with the composition of $Co_{46}Mn_{27}Si_{27}$ was used for preparing the CMS layers. The composition of the deposited CMS layers analyzed by inductively coupled plasma (ICP) analysis was $Co_{50.3}Mn_{23.4}Si_{26.4}$. The structure of the films was examined by X-ray diffraction (XRD) and cross sectional transmission electron microscopy (TEM).

Multilayer films were deposited on a MgO(001) single crystalline substrate at RT with the stacking structure of Cr(10)/Ag(200)/Cr(10)/CMS(20)/Cu(4)/CMS (5)/$Fe_{25}Co_{75}$ (2)/$Ir_{22}Mn_{78}$ (10)/Ru(5), where the numbers in the parentheses indicate the thickness in nm. After depositing the lower CMS layer of 20 nm, the film was annealed at 673 K to improve the chemical ordering of the structure. The film was cooled to RT and then the remaining layers were deposited at RT. The film was microfabricated by combining electron beam lithography and Ar ion etching for measuring MR in the CPP geometry. The actual area measured by atomic force microscopy was $0.95\times0.51$ $(\mu m)^2$. After fabricating and forming electrodes for MR measurements, the samples were annealed at 573 K in a magnetic field of 5 kOe to introduce exchange biasing by the antiferromagnetic $Ir_{22}Mn_{78}$ layer. The magnetic field was applied parallel to the [110] direction of CMS in the plane, which is the easy axis of the CMS. The MR curve was measured at 6 K and RT by the 4-probe measurement method.

Figure 1 shows the x-ray diffraction (XRD) pattern of Cr(10)/Ag(200)/Cr(10)/CMS(20)/Ru(5) layers on the MgO(100) substrate obtained by the conventional $2\theta$-$\theta$ scan. The film was annealed at 673K after depositing the CMS layer. The results indicate that all the layers of Cr, Ag and CMS were grown with the [001] orientation, suggesting the epitaxial growth of the layers. The {002} peak indicates the *B2*-type ordering of CMS. The inset of Fig. 1 shows the Φ scan of the CMS (111) reflections. The peaks with the four-fold symmetry indicate the epitaxial growth with the $L2_1$ structure.

Figure 2(a) shows a two beam ($\boldsymbol{g}$=$(004)_{Ag}$) bright-field TEM image of the sample that was used for CPP-MR measurements, which shows relatively flat interfaces between the layers. Nano-beam electron diffractions taken from the [110] axis of the CMS layers indicate that the bottom CMS layer that was *in situ* annealed at 673 K has the $L2_1$ ordered structure (Fig. 2(d)). On the other hand, the top CMS layer was found to have the B2 structure (Fig. 2(c)), which was *ex situ* annealed at 573 K in magnetic field. The diffraction patterns from both the CMS and the Cu layers showed the epitaxial relationship of CMS(100)[110]//Cu(100)[100]. In addition, a small local lattice rotation of the Cu [110] axis to the out-of-plane direction with respect to the CMS [100] axis was found with an angle of about 5 degrees. Figure 2(b) shows the high resolution (HR) TEM image near the CMS/Cu/CMS interfaces. The lattice rotation is also observed in this image. A lot of defects including lattice distortions and dislocations are observed in the Cu spacer layer. Note that the top CMS layer grew epitaxially on the Cu layer despite high density of the lattice defects in the Cu layer and that the Cu has a relatively large lattice mismatch of -9.7% with respect to the CMS. Since the Cu layer receives a tensile stress derived from the mismatch from the bottom and top CMS layer, the defects might be slip zones or twins in the Cu (111) plane. The deformation of the Cu layer, as well as the lattice rotation observed above, seems to occur for relaxing the large mismatch at the interface. It is considered that the plastic deformation within the Cu spacer layer is considered to accommodate the mismatch, thereby enabling the epitaxial growth of the top CMS layer. The detailed structural analysis will be reported elsewhere.

Figure 3(a) shows the magnetic field dependence of the resistance-area product of the CPP-GMR spin valve at RT. The resistance in the parallel ($R_P$) and antiparallel ($R_{AP}$) alignments of the magnetization were obtained. At RT, the resistance-area product was $R_PA$ = 0.165 $\Omega(\mu m)^2$. The resistance change-area product was $\Delta RA = R_{AP}A - R_PA$ = 14.2 m$\Omega(\mu m)^2$, and the MR ratio was 8.6%. For comparison, we also examined the sample without Ag buffer layer, having the layer structure of Cr(30)/CMS(20)/Cu(4)/CMS(5)/$Fe_{25}Co_{75}$(2)/$Ir_{22}Mn_{78}$(10)Ru(5) on MgO(001) and treated in the same annealing conditions. The results are $R_PA$ = 1.08 $\Omega(\mu m)^2$, $\Delta RA$ =15.2 m$\Omega(\mu m)^2$ and the MR ratio of 1.4 %. Yakushiji *et al.* [3)] reported the values $R_PA$ = 0.79 $\Omega(\mu m)^2$, $\Delta RA$ =19 m$\Omega(\mu m)^2$, and the MR ratio of 2.4 % by using the CMS/Cr/CMS structure on the Cr buffer layer. Thus, the present work produced a smaller $R_PA$ because of the Ag layer inserted in the buffer layer. This is the main reason why the substantially larger MR ratio was obtained despite its slightly lower $\Delta RA$ compared to the previous work. The epitaxial growth of the both CMS layers possibly leads to the high MR ratio due to the enhancement of $\Delta_1$ band electrons contribution to spin polarized transport. It is also likely that the interface resistance was minimized by the epitaxial growth of both CMS layers.

As shown in Fig. 3 (b), we obtained the much larger MR ratio of 30.7% and $\Delta RA$ = 35.2 m$\Omega$($\mu$m)$^2$ at 6K. These are the highest values reported thus far for CPP-GMR devices. The origin of the large CPP-GMR at low temperatures is obviously the half metallicity of CMS. As reported by Sakuraba *et al.*[10] from their TMR measurements of the CMS/Al-O/CMS MTJ, one drawback of CMS is its large degradation of half-metallicity at RT. A similar tendency was observed in our CPP-GMR using CMS, i.e., the high MR ratio of 30.7% measured at 6K decreases to 8.6% at room temperature. One possible explanation for the decrease is the imperfectness of the half-metallicity caused by the disordering of the $L2_1$ structure of the CMS electrodes. The disorder may lead to a decrease in the energy gap in the minority spin band. Then, the conduction through the minority spin band would increase by the thermal excitation at elevated temperatures, resulting in a reduction of MR. Therefore, the improvement of the $L2_1$ ordering of the CMS layers would lead to a further improvement of the GMR ratios.

In summary, spin-valve structures with epitaxial trilayers of CMS/Cu/CMS were prepared on MgO (001) single crystal substrates with the Cr/Ag/Cr buffer layers. Structural studies have shown the (001) oriented epitaxial growth up to the upper CMS layer despite the large lattice mismatch between Cu and CMS. The bottom CMS layer annealed at 673 K has the $L2_1$ structure, while the top CMS layer was in the B2 structure. The large CPP-GMR ratio of 8.6% and $\Delta RA$=14.2 m$\Omega$ ($\mu$m)$^2$ with the low resistance area product $R_PA$ = 0.165 $\Omega$ ($\mu$m)$^2$ were attained at RT. The MR ratio increased with decreasing temperature to 30.7% at 6 K, which is the highest MR value reported for CPP-GMR. The MR characteristics are expected to show further improvement by increasing the degree of the $L2_1$ orders of the CMS layers.

KK and TMN thank the National Institute for Materials Science for providing a NIMS junior research assistantship. This work was partly supported by a Grant-in-Aid for Scientific Research (B) 20360322, a Grant-in-Aid for Scientific Research in Priority Area "Creation and control of spin current" 19048029, and the World Premier International Research Center Initiative (WPI Initiative) on Materials Nanoarchitronics, MEXT, Japan.

Fig. 1: X-ray diffraction (XRD) pattern of the 2θ-θ scan for the Cr(10)/Ag(200)/Cr(10)/CMS(20)/Ru(5) stack. The sample was annealed at 673 K after depositing the CMS layer. The inset shows the Φ scan for the (1 1 1) reflections.

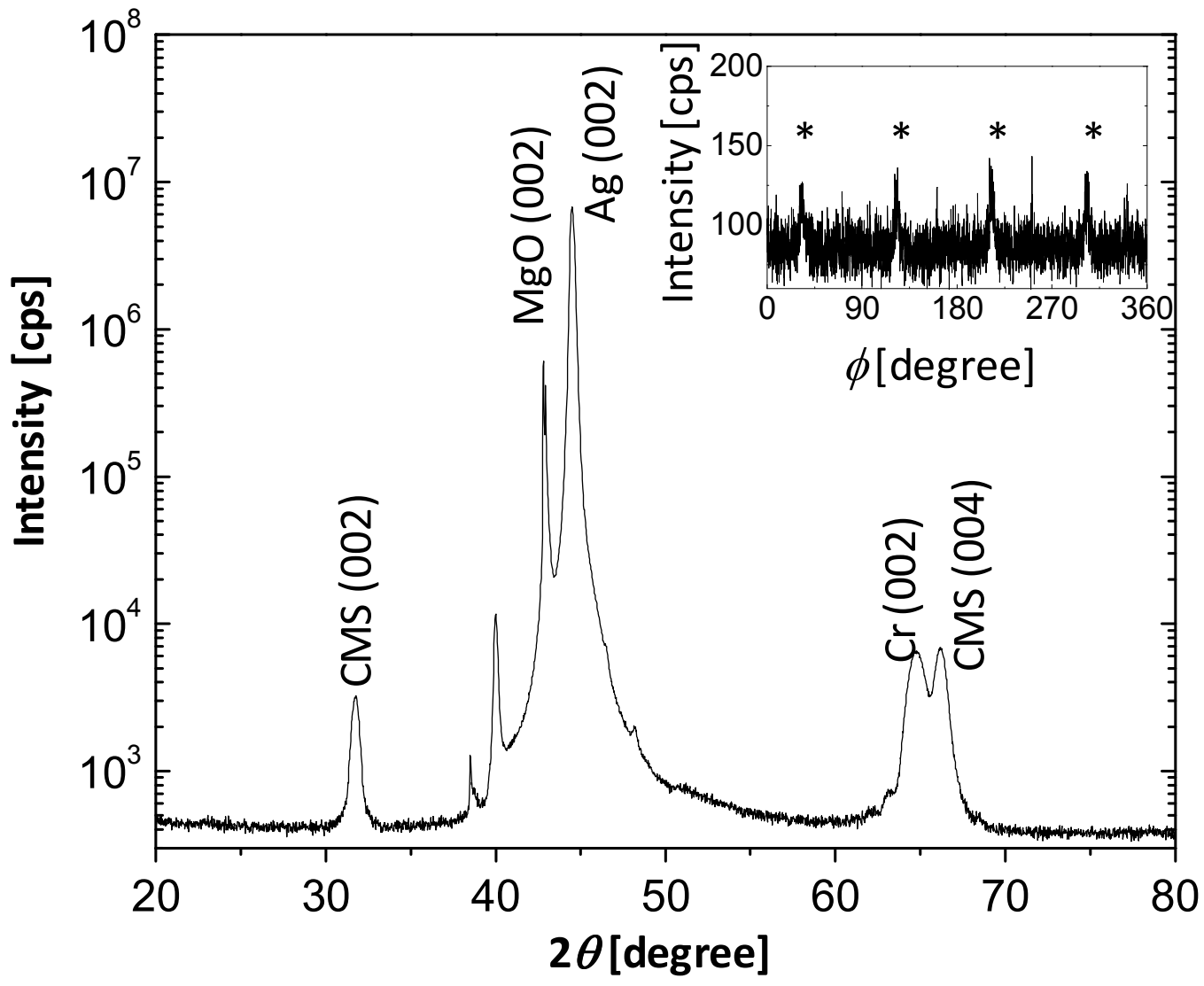

Fig. 2: Cross sectional transmission electron micrographs of the spin valve; (a) two beam condition bright field image, (b) high resolution image, (c) nano-beam diffraction pattern taken from the top CMS layer, and (d) nano-beam diffraction pattern taken from the bottom CMS layer. The white auxiliary lines show the lattice rotation of about 5º at the CMS/Cu epitaxial interface. The electron beam is parallel to the [110] direction of CMS.

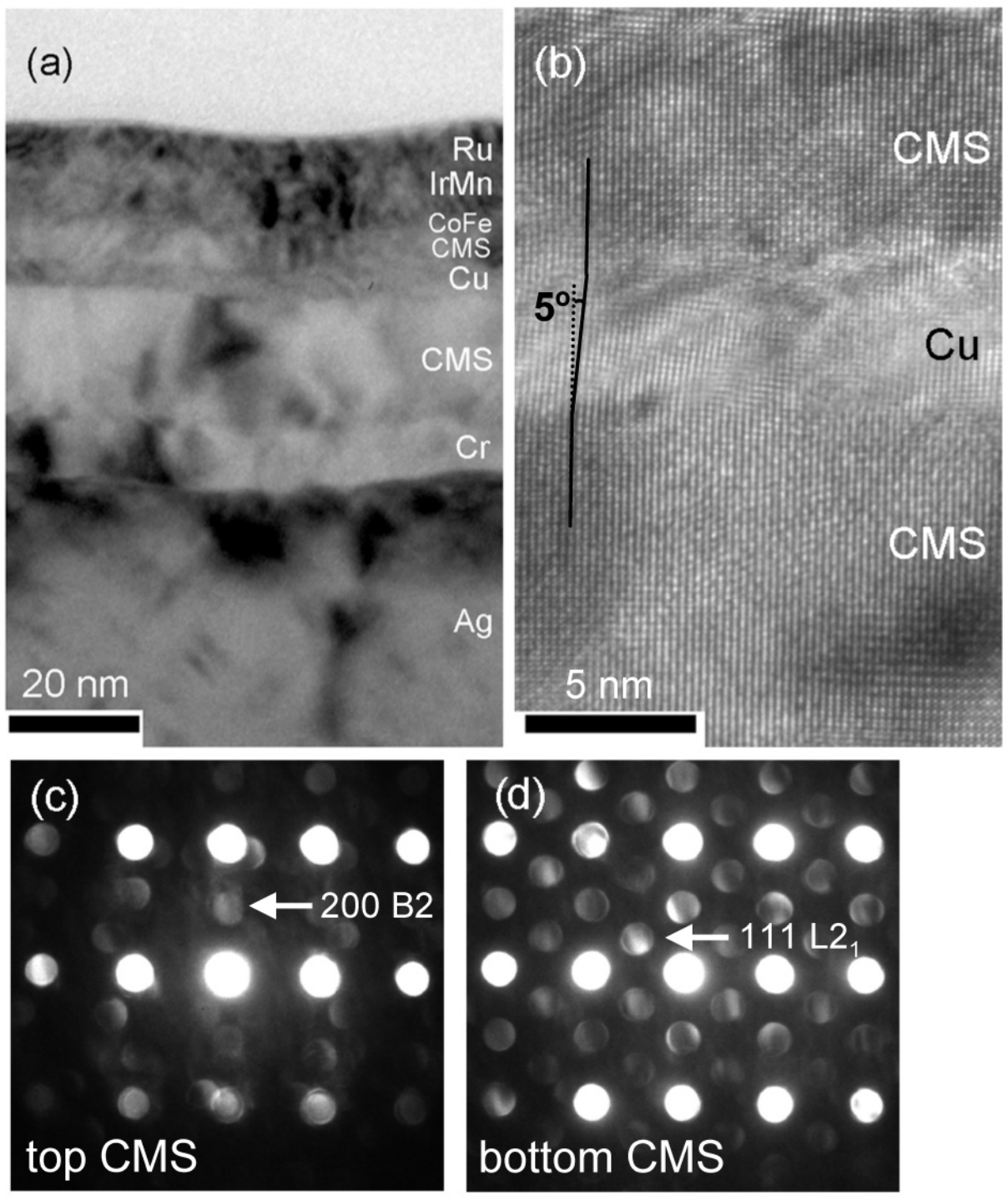

Fig 3: Magnetic field dependence of the resistance-area product of the MgO(001)/Cr/Ag/Cr/CMS/Cu/CMS /$Co_{75}Fe_{25}$/$Ir_{22}Mn_{78}$/Ru spin valve at (a) RT and (b) 6 K.

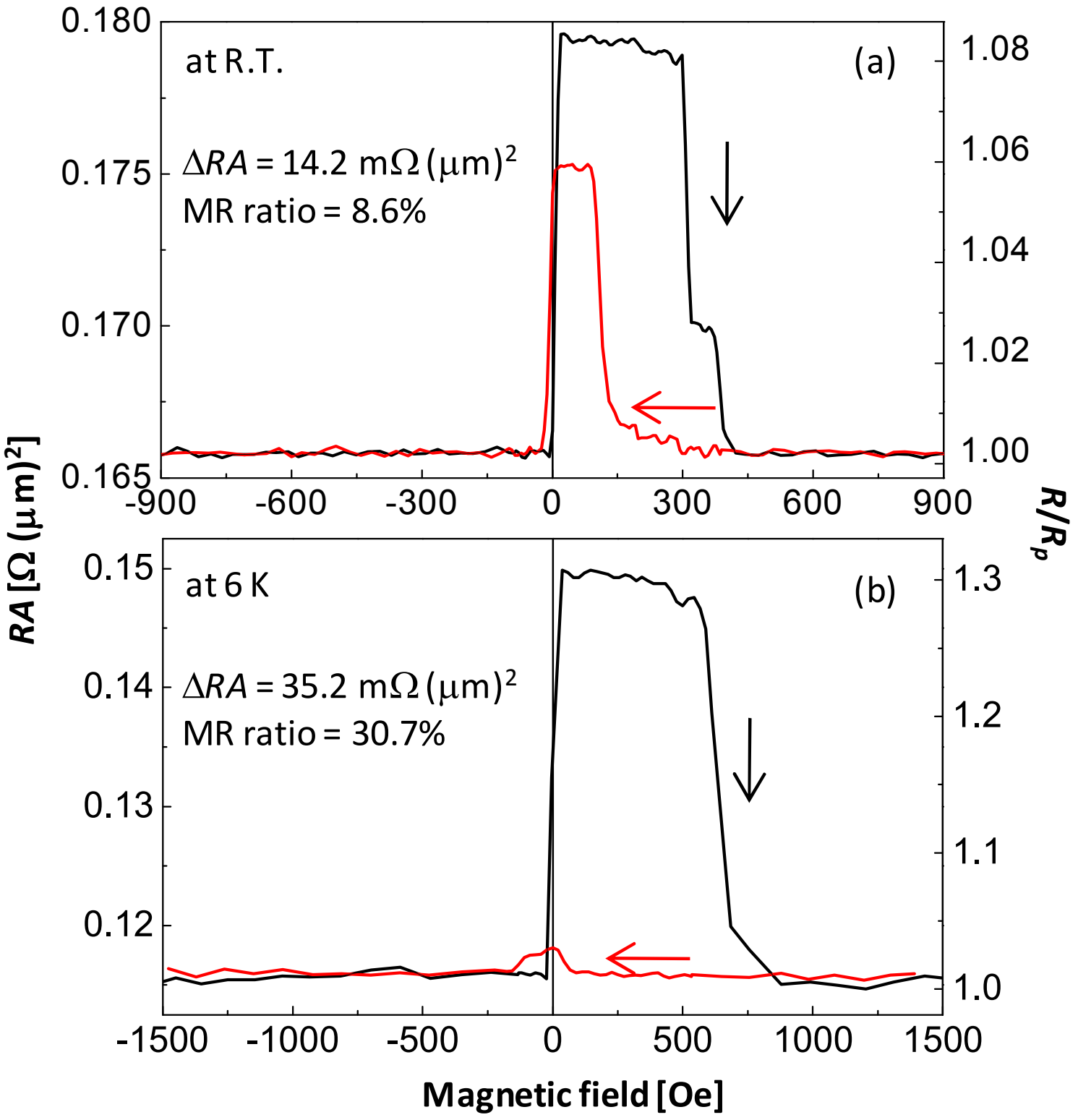